\documentclass[12pt,runningheads]{llncs}
\usepackage{graphicx}
\usepackage{float}
\usepackage[margin=30 mm]{geometry} 
\usepackage{arydshln}
\usepackage[misc]{ifsym}

\title{\begin{huge}
\textbf{Multimodal PET/CT Tumour Segmentation and Prediction of Progression-Free Survival using a Full-Scale UNet with Attention} 
\end{huge}}

\author{Emmanuelle Bourigault\orcidID{0000-0002-5270-6836}\inst{1}  \and Daniel R. McGowan \orcidID{0000-0002-6880-5687}\inst{2} \and Abolfazl Mehranian \orcidID{0000-0003-4584-4453}\inst{3} \and Bart\l omiej W. Papie\.z \orcidID{0000-0002-8432-2511}\inst{4,5}}

\institute{
	Department of Engineering, Doctoral Training Centre, University of Oxford, Oxford, UK
	\and
	Department of Oncology, University of Oxford, Oxford, UK
	\and
	GE Healthcare, Oxford, UK
	\and
	Big Data Institute, Li Ka Shing Centre for Health Information and Discovery, University of Oxford, Oxford, UK
	\and 
	Nuffield Department of Population Health, University of Oxford, Oxford, UK
}
\authorrunning{Bourigault E. et al.}
\titlerunning{Multimodal PET/CT tumour segmentation and patient survival prediction}

\begin{document}

    \maketitle
    \begin{abstract}
    \normalsize
    Segmentation of head and neck (H\&N) tumours and prediction of patient outcome are crucial for patient's disease diagnosis and treatment monitoring. Current developments of robust deep learning models are hindered by the lack of large multi-centre, multi-modal data with quality annotations. The MICCAI 2021 HEad and neCK TumOR (HECKTOR) segmentation and outcome prediction challenge creates a platform for comparing segmentation methods of the primary gross target volume on fluoro-deoxyglucose  (FDG)-PET and Computed Tomography images and prediction of progression-free survival in H\&N oropharyngeal cancer.
    For the segmentation task, we proposed a new network based on an encoder-decoder architecture with full inter- and intra-skip connections to take advantage of low-level and high-level semantics at full scales. 
    Additionally, we used Conditional Random Fields as a post-processing step to refine the predicted segmentation maps. 
    We trained multiple neural networks for tumor volume segmentation, and these segmentations were ensembled achieving an average Dice Similarity Coefficient of 0.75 in cross-validation, and 0.76 on the challenge testing data set.
    For prediction of patient progression free survival task, we propose a Cox proportional hazard regression combining clinical, radiomic, and deep learning features. 
    Our survival prediction model achieved a concordance index of 0.82 in cross-validation, and 0.62 on the challenge testing data set.
    
    \end{abstract}
    
    \section{Introduction}
    \indent Head and Neck (H\&N) tumour is the fifth most prevalent cancer worldwide. 
    Improving the accuracy and efficiency of disease diagnosis and treatment is the rationale behind the developments of computer-aided systems in medical imaging \cite{ref_article1,ref_article2}.
    However, obtaining manual segmentations, which can be used for diagnosing and treatment purposes, is time consuming and suffers from intra- and inter-observer biases. 
    Furthermore, segmentation of H\&N tumours is a challenging task compared to other parts of the body as the tumour displays similar intensity values to the adjacent tissues making it non distinguishable to the human eye in Computed Tomography (CT) images. Previous attempts at developing deep learning models to segment head and neck tumours suffered from a relatively high number of false positives \cite{ref_article3,ref_article4}.
    Currently, in the normal clinical pathway, a combination of Positron Emission Tomography (PET) and CT images plays a key role in the diagnosis of H\&N tumors.
    This multi-modal approach has dual benefits: the metabolic information is provided by PET and anatomical information is available in CT.
    Furthermore, accurate segmentation of H\&N tumors could also be used in automating pipelines for extraction of quantitative imaging features (e.g. radiomics) in prediction of patient survival.\\
    \indent The 3D UNet \cite{ref_article5} is one of the most widely employed encoder-decoder architecture for medical segmentation inspired by Fully Convolutional Networks \cite{ref_article6}. 
    Promising results have been obtained using 3D UNet based architecture and attention mechanisms with early fusion of PET/CT images~\cite{ref_article7}. 
    While the performance of the model proposed in~\cite{ref_article8} was significantly improved when compared to a baseline 3D UNet, a number of false positives was reported where the model was not only segmenting the primary tumour but also other isolated areas such as the soft palate due to tracer overactivity in that region. \\
    \indent In this paper, we propose to segment 3D H\&N tumor volume from multimodal PET/CT imaging using a full scale 3D UNet3+ architecture \cite{ref_article9} with attention mechanism.
    Our model, NormResSE-UNet3+, is trained with a hybrid loss function of Log Cosh Dice~\cite{ref_article10} and Focal loss~\cite{ref_article11}. 
    The segmentation maps, predicted by our model, are further refined by Conditional Random Fields post-processing~\cite{ref_article12,ref_article13} to reduce number of false positives and to improve tumour boundary segmentation. 
    For the progression free survival prediction task, we propose a Cox proportional hazard regression model using a combination of clinical, radiomic, and deep learning features from PET/CT images.\\
    \indent The paper is organised as follows. 
    Section~\ref{data_methods} outlines the data set and pre-processing steps (Section \ref{sec_data}), the methods used for 3D H\&N tumor segmentation and for prediction of progression free survival tasks (Section \ref{sec_pfs}), and the evaluation criteria (Section~\ref{sec_evaluation}).
    The experimental set-up and results are described in Section~\ref{results}. 
    Finally, the discussion and conclusion are found in Section~\ref{conclusion}.
    \newpage
    
    \section{Methods and Data}
    \label{data_methods}
    
    \subsection{Data} 
    \label{sec_data}
    
    \subsubsection{Data for Segmentation.}
    PET and CT images used in this challenge were provided by the organisers of the HECKTOR challenge at the 24th International Conference on Medical Image Computing and Computer Assisted Intervention (MICCAI)\cite{ref_article1}.
    The total number of training cases is 224 from 5 centers: CHGJ, CHMR, CHUS, CHUP, and CHUM. 
    The ground-truth annotations are provided by expert clinicians for primary gross tumor volume.
    For testing, additional 101 cases from two centers, namely CHUP and CHUV, are provided. However, no expert annotations are available to the participants.
    
    \subsubsection{Data Preprocessing for Segmentation Task.} 
    For segmentation task, we used trilinear interpolation to resample PET and CT images. Bounding boxes of $144$x$144$x$144$ voxels were provided by the organizers and used for patch extraction. 
    PET intensities (given in standard uptake value) were normalised with Z-score, while CT intensities (given in Hounsfield unit) were clipped to the range~$[-1, 1]$.

    \subsubsection{Data for Progression Free Survival.}
    Patient clinical data are provided for prediction of progression free survival in days. 
    The covariates (a combination of categorical and continuous variables) are as follows: center ID, age, gender, TNM 7/8th edition staging and clinical stage, tobacco and alcohol consumption, performance status, HPV status, treatment (radiotherapy only, or chemoradiotherapy). 
    Dummy variables were used to encode the categorical variables i.e. the ones mentioned above except for age (continuous).
    Among the 224 patients for training, the median age was 63 years (range: 34-90 years) with progression event occurred in 56 patients and an average progression survival of 1218 days (range: 160-3067 days).
    The testing cohort comprised 129 patients with a median age of 61 years (range: 40-84 years).
    
    \begin{figure}[tbh]
    \includegraphics[width=\textwidth]{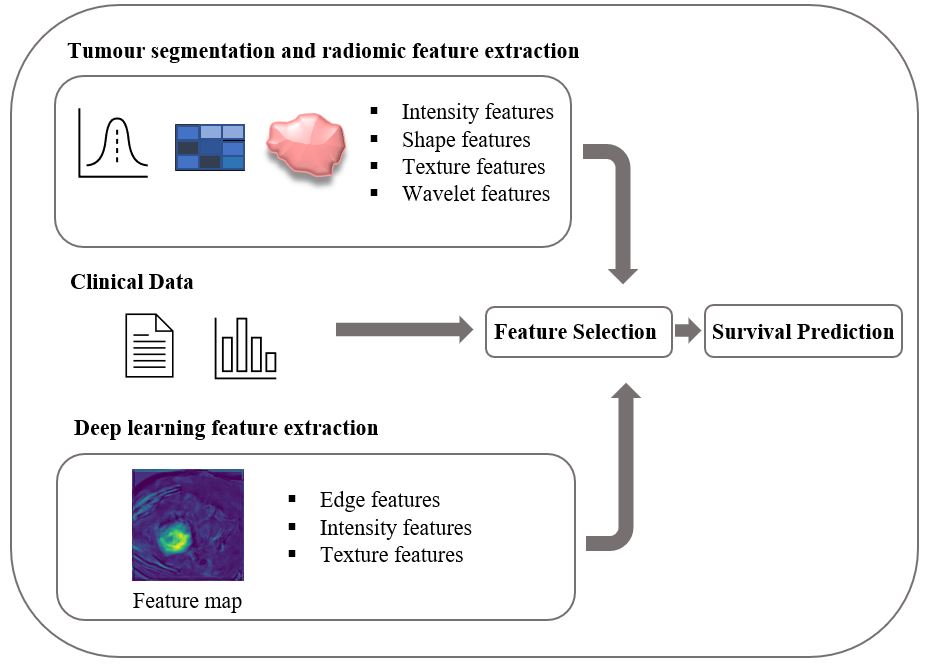}
    \centering
    \caption{The pipeline used for the Progression Free Survival task. It consists of three steps following clinical data collection, images acquisition, and pre-processing, then feature extraction, feature selection, and statistical survival prediction. 
    }
    \label{fig 1:pipeline survival}
    \end{figure}

    \subsubsection{Data Preprocessing for Progression Free Survival Task.} 
    For the prediction of progression free survival, we used multiple imputation for missing values. The multiple imputation models each feature, which contains missing values, as a function of the other features. 
    Then, it uses this estimate in a round-robin fashion for imputing the missing values. At each iteration, a feature is designated as output $\mathbf{y}$ and the other features are treated as inputs $\mathbf{X}$. 
    A regressor is fit on $\mathbf{(X, y)}$ on the known $\mathbf{y}$ used to predict the missing values of $\mathbf{y}$. This process is repeated for each feature and for ten imputation rounds. 
    Feature selection was performed using Lasso regression with 5-fold cross-validation using a combination of features i.e. clinical features, radiomic features, and features extracted from 3D UNet (see Fig.~\ref{fig 1:pipeline survival}).
    The correlation between those features was evaluated with Spearman’s correlation coefficient in order to assess potential redundancy. 
    A threshold of 0.80 was set to filter out highly correlated features. 
    The feature selection process reduced the number of features from 275 to
    70 most relevant features. In particular, we kept 7 clinical features (i.e. Age, Chemotherapy, Tstage, Nstage, TNMgroup, TNM edition, and HPV status), 14 radiomics (5 intensity based histogram features, 3 shape features, and 6 texture features from metrics: Gray Level Co-occurrence Matrix (GLCM),  Gray Level Run Length Matrix (GLRLM), Gray Level Size Zone (GLSZM)), and 49 deep learning features.
    Radiomic features were extracted from PET and CT images using the pyradiomics package. Convolutional Neural Networks (CNNs) by using stacks of filtering layers together with pooling and activation layers, are becoming increasingly popular in the field of radiomics ~\cite{ref_article14}. This can be explained by the fact that CNNs do not require prior knowledge and kernels are learned automatically as opposed to hand-crafted features. In this study,deep learning features were extracted at the 5th convolutional layer of our model by averaging feature maps. A vector was created for each feature map, then concatenated to form a single vector of deep learning features. The power of CNNs is their ability to automatically learn multiple filters in parallel thus extracting low and high-level features such as edges, intensity but also texture. Each filter captures different characteristics of the image ultimately enabling CNNs to capture relevant edge, intensity and texture features ~\cite{ref_article15} (see a schematic overview of the survival pipeline in Fig.\ref{fig 1:pipeline survival}).

    \subsection{Models Description}
    \subsubsection{Models for Segmentation Task.} 
    \label{sec_seg}
\begin{figure}[tbh]
    \centering
    \includegraphics[width=\textwidth]{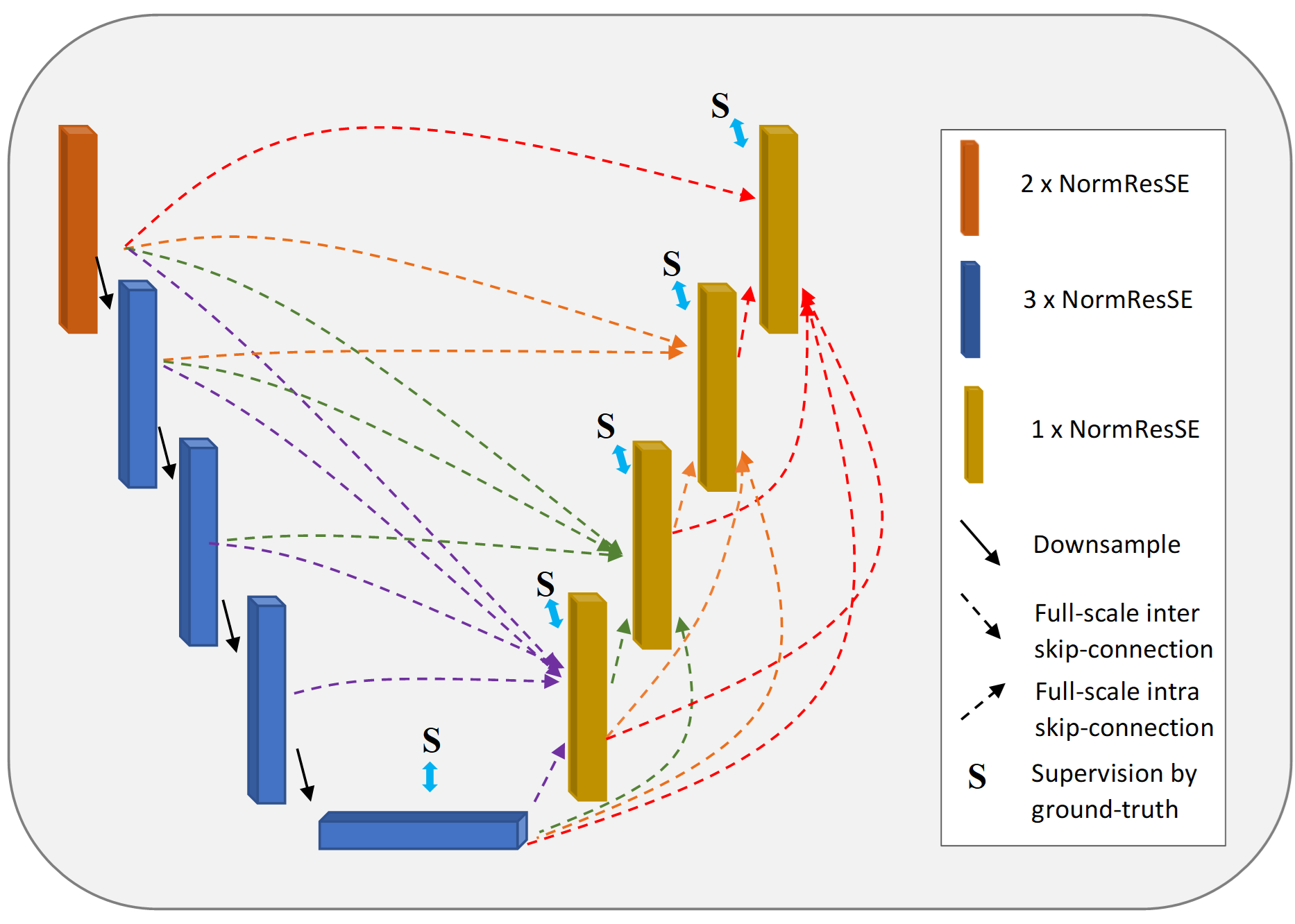}
    \caption{Architecture of NormResSE-UNet3+ inspired from ~\cite{ref_article8,ref_article9}. The input is a 2 × 144 × 144 × 144 tensor consisting of concatenated PET and CT images. 
    The encoder is made of residual squeeze-and-excitation blocks whose first block has 24 filters following~\cite{ref_article17}. The output of the encoder has dimension 384x3x3x3. The decoder path contains full-scale inter- and intra-skip connections as well as a supervision module by ground-truth. The decoder outputs one channel with same size as the input (1 x 144 x 144 x 144).} \label{fig1:architecture NormResSE UNet3+}
    \end{figure}

    The UNet model \cite{ref_article5}, an encoder-decoder architecture, is one of the most widely employed segmentation models in medical imaging.
    Skip connections are used to couple high-level feature maps obtained by the decoder and corresponding low-level feature maps by the encoder. 
    UNet++ is an extension of UNet that introduce nested and dense skip connections to reduce the merging of dissimilar features from plain skip connections in UNet~\cite{ref_article16}. 
    However, since UNet++ still fails to capture relevant information from full scales and to recover lost information in down- and up-sampling, \cite{ref_article9} proposed UNet3+ to take full advantage of multi-scale features. 
    The design of inter- and intra-connection between the encoder and decoder pathways at full scale enables us to explore both fine and coarse level details (see a schematic overview in Fig.~\ref{fig1:architecture NormResSE UNet3+}). 
    Low level details contain information about the spatial and boundary information of the tumour, while high-level details encode information about the location of tumour. The integration of deep supervision in the decoder pathway is used to reduce false positives. 
    To further reduce false positives and improve segmentation, we make use of attention mechanisms achieving state-of-the-art segmentation results~\cite{ref_article17}. 
    In particular, we use 3D normalised squeeze-and-excitation residual blocks proposed by~\cite{ref_article8} and evaluated on PET/CT H\&N dataset from MICCAI 2020 challenge~\cite{ref_article7}.
    
    \subsubsection{Loss function for Segmentation Task.} 
    The Dice Coefficient is a widely used loss function for segmentation tasks, and is defined as follows:
    \begin{equation}
    DiceLoss(y,\hat{p})=1-
    \frac{2\sum_{i}^{N}y_{i}\hat{p}_{i} + 1}{\sum_{i}^{N}y_{i} + \sum_{i}^{N}\hat{p}_{i} + 1}
    \label{eq:Dice Loss}
    \end{equation}
    In addition, 1 is added in the numerator and denominator to ensure that
    the function is not undefined in cases when y = $\hat{p}$ = 0, i.e. the tumour is not present. 
  
    The Focal loss~\cite{ref_article11} is a variation of the Cross-Entropy loss. 
    The Focal loss is well-suited for imbalance problems as it down-weights easy examples to focus on hard ones, and is defined as follows:
    \begin{equation}
    FocalLoss(y,\hat{p}) = \mathbf{-}\frac{1}{N}\sum_{i}^{N}y_{i}(1-\hat{p}_{i})^{\gamma }ln(\hat{p}_{i})
    \end{equation}
    where $\gamma$ in the modulating factor is optimised at 2.
    The Log-Cosh is also popular for smoothing the curve in regression problems~\cite{ref_article10}.

    For the data in the HECKTOR challenge, we tested the abovementioned loss functions and their combinations,  and in our best performing model, we used a hybrid loss function, the Cosh Log Dice loss combined with the Focal loss defined as follows:
    \begin{equation}
    LogCoshDiceFocalLoss = log(cosh(DiceLoss)) + FocalLoss
    \label{eq:LogCoshDiceFocal}
    \end{equation}
    
    \subsubsection{Refining segmentation maps.}
    We used 3D Conditional Random Fields (CRF) to refine segmentation maps ~\cite{ref_article12,ref_article13}. 
    The segmentation output produced by CNNs tend to be too smooth because of neighbouring voxels sharing spatial information.
    CRF is a graphical model that captures contextual, shape and region connectivity information thus becoming a popular refinement procedure to improve segmentation performance, for example, \cite{ref_article18} used CRF to refine the segmentation outputs as a post-processing step.

    \subsubsection{Models for Progression Free Survival Task.} 
    \label{sec_pfs}
  
    Cox proportional hazard (CoxPH) regression model is the most commonly used hazard model in the medical field because it effectively deals with censoring. 
    Random Survival Forest is also a popular model for survival time prediction working better for big sample sizes \cite{ref_article19,ref_article20}. It builds an ensemble of trees on different bootstrap samples of the training data before aggregating the predictions.
    DeepSurv~\cite{ref_article21} showed improvements over traditional CoxPH model as it better captures the complex relationship between a patient’s features and effectiveness of different treatments.  
    DeepSurv is a Cox proportional hazards deep neural network, which estimates the individuals' effect based on parametrized weights of the neural network.
    The architecture is a multi-layer perceptron configurable with the number of hidden layers. In this study, we used  32 hidden layers which are fully-connected nonlinear activation layers. 
    Dropout layers are added to reduce over-fitting. The output layer of DeepSurv has a single node with linear activation function to give estimates of log-risk hazard. Compared to traditional Cox regression, which is optimized with the Cox partial likelihood, DeepSurv uses the negative log partial likelihood with the addition of a regularization term. DeepSurv achieved state-of-the-art results for cancer prognosis prediction with concordance index close or higher than 0.8 ~\cite{ref_article22,ref_article23}.

    \subsection{Evaluation Metrics}
    \label{sec_evaluation}
    \subsubsection{Evaluation Metrics for Segmentation Task.}
    \indent The Dice Similarity Coefficient (DSC) is a region-based measure to evaluate the overlap between the prediction (P) and the ground truth (G). 
    DSC is given as follows:
    \begin{equation}
     DSC(P, G) =  \frac {2 |P \cap G|} {|P| + |G|}  
    \label{eq:Dice score}
    \end{equation}
    The DSC ranges between 0 and 1, with a larger DSC denoting better performance.\\
    The average Hausdorff distance (HD) between the voxel sets of ground truth and segmentation is defined as:
    
    \begin{equation}
    Average HD = \frac{1}{2} (\frac{GtoP} {G} + \frac{PtoG} {P})
    \label{eq:Hausdorff Distance}
    \end{equation}
    where GtoP is the directed average HD from the ground-truth to the segmentation, PtoG is the directed average HD from the segmentation to the ground truth, G is the number of voxels in the ground truth, and P is the number of voxels in the segmentation. The 95th percentile of the distances between voxel sets of ground truth and segmentation (HD95) is used in this work to reduce the impact of outliers.

    \subsubsection{Evaluation Metrics for Survival Task.}
    Harrell's concordance index (C-index) is the most widely used measure of goodness-of-fit in survival models.  It is defined as the ratio of  correctly ordered (concordant) pairs divided by the total number of possible evaluation pairs. The C-index is used in this study to evaluate survival prediction outcome as it takes into account censoring. The C-index quantifies how well an estimated risk score is able to discriminate among subjects who develop an event from those who do not. 
    In this work, the event of interest is progression. The C-index ranges between 0 and 1 with 1 denoting perfect predicted risk.
    
    Code available at: \url{https://github.com/EmmanuelleB985/Head-and-Neck-Tumour-Segmentation-and-Prediction-of-Patient-Survival}
    \newpage
    \section{Results}
    \label{results}
    \subsection{Segmentation Task}
\begin{figure}[H]
    \centering
    \includegraphics[width=\textwidth]{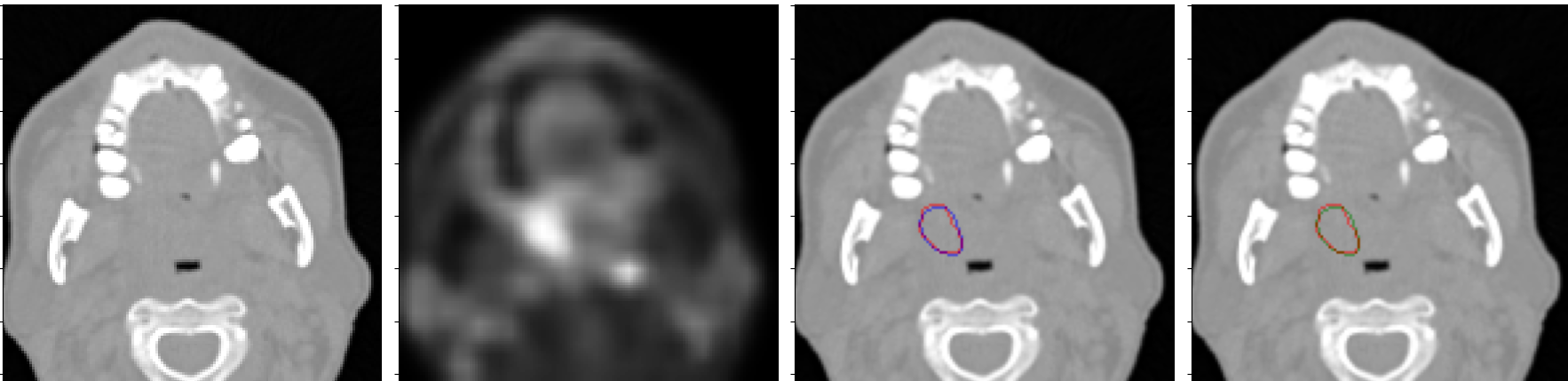}
    \includegraphics[width=\textwidth]{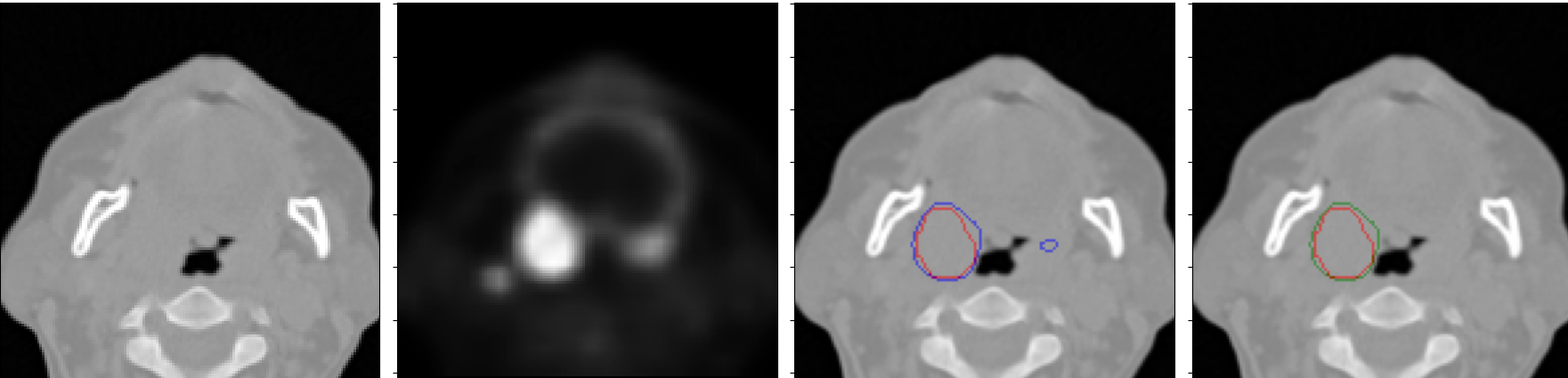}
    \includegraphics[width=\textwidth]{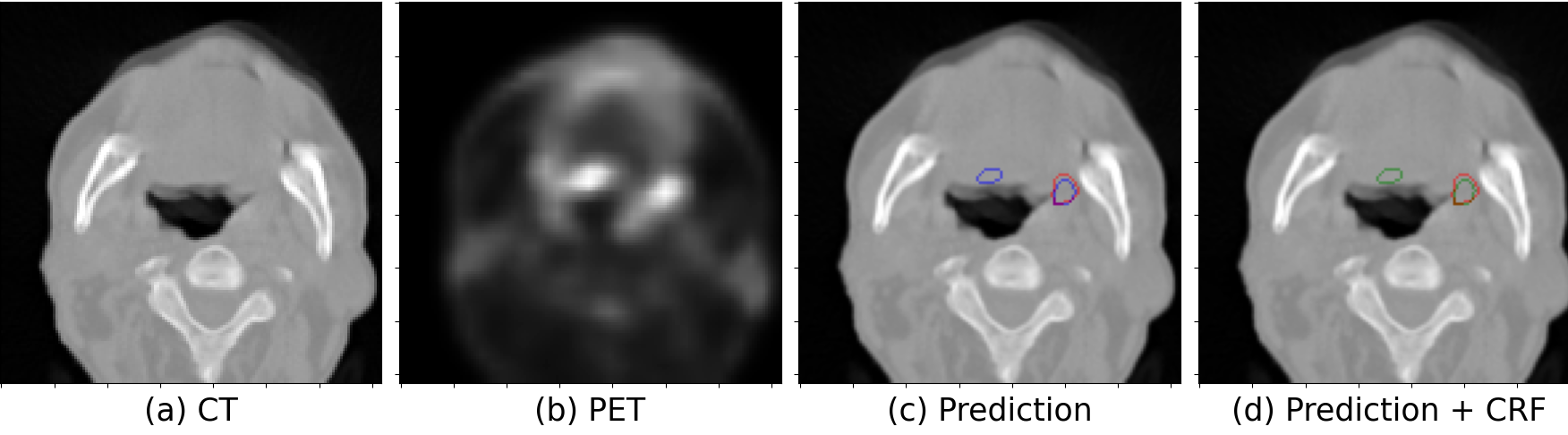}
    \caption{Examples of segmentation predictions. 
    From left to right, the axial view of (a) CT image, (b) PET image, and 
    (c) CT overlaid with a segmentation produced by our model (blue) and ground-truth tumour annotation (red), and
    (d) CT overlaid with segmentation refined by the CRF (green), and ground-truth tumour annotation (red).
    Two top rows show the examples of the good quality segmentations, while the third row shows an example of the partially correct segmentation, where false positives were not removed by the post-processing. 
    } 
    \label{CHUP017_fp.png}
    \end{figure}

    The model was trained on 2 NVIDIA A100 GPUs for 1000 epochs. The optimizer used is Adam (0.9,0.999). 
    The scheduler is cosine annealing with warm restarts with the input learning rate value of $10^{-2}$, and reducing the learning rate every 25 epochs. 
    A batch size of 2 was used for training and validation. Data augmentation, namely random flipping and random rotation, is used during training to reduce over-fitting.
    Lifelines and Pycox packages were used for all statistical analyses.
    \\We trained the 3D NormResSE-Unet3+ on a leave-out one center, and we performed model ensembling by averaging the predictions on the test set of the 5 models trained (see Fig.~\ref{CHUP017_fp.png} and Tab.~\ref{table:seg}). 
    An example of a good quality segmentation map predicted by our model is shown in Fig.~\ref{CHUP017_fp.png} (the first row). 
    An example of the predicted segmentation map, which benefited from the CRF post-processing to reduce false positives is shown in Fig.~\ref{CHUP017_fp.png} (the second row).
    An example of failure of our pipeline to discard false positives from true primary tumour is shown in Fig.~\ref{CHUP017_fp.png} (the third row).
    The quantitative results are summarised in Tab.~\ref{table:seg}. 
    For each fold, the segmentation results are presented in terms of DSC. 
    We obtained an average DSC of 0.753 and an average Hausdorff Distance at 95\% (HD95) of 3.28 with post-processing and ensembling techniques. 
    On the test set provided by HECKTOR2021, our model achieved an average DSC of 0.7595 and HD95 of 3.27, showing good generalisability. 
    \newpage

    \begin{table}[tbh]
    \caption{Quantitative results for Segmentation Task. The multiple neural networks for H\&N tumor segmentation were trained, and these segmentations were ensembled achieving an average DSC of 0.75 and an average HD95 of 3.28 in cross-validation. } 
    \centering 
    \begin{tabular}{l c c c c c} 
    \hline 
    Cross-validation fold & \multicolumn{2}{c}{NormResSE-UNet3+} &
    &\multicolumn{2}{c}{NormResSE-UNet3+ + CRF} \\
    & DSC & HD & & DSC & HD95 \\
    \hline\hline 
    Fold 1 & 0.792 & 3.18 && 0.822 & 3.11\\ 
    Fold 2 & 0.693 & 3.43 && 0.702 & 3.41\\
    Fold 3 & 0.728 & 3.32 && 0.749 & 3.29\\
    Fold 4 & 0.736 & 3.31 && 0.738 & 3.30\\
    Fold 5 & 0.742 & 3.29 && 0.756 & 3.28\\\hdashline
    Ensemble & 0.738 & 3.30 && 0.753 & 3.28\\ [1ex] 
    \hline 
    \end{tabular}
    \label{table:seg} 
    \end{table}

    \subsection{Survival Task}
    
    We trained three models: CoxPH regression, Random Survival Forest, and DeepSurv on 5-fold cross-validation splits. 
    Each of the above models were trained with different configurations of clinical, PET/CT radiomic and deep learning features. 
    CoxPH regression was trained with a combination of clinical, CT radiomics, and deep learning features, and achieved the best significant c-index of 0.82 (p-value $<$ 0.05) using a corrected paired two-tailed t-test at the 5\%  significance level to compare each pair of models \cite{ref_article24} (see Tab.~\ref{table:surv}). 
    The second best c-index of 0.75 was obtained with CoxPH regression trained with clinical and deep learning features (see Tab.~\ref{table:surv}). 
    
    The performance on the test set provided by HECKTOR 2021 with the CoxPH regression using clinical, CT radiomics and deep learning features was significantly lower i.e. 0.62 suggesting over-fitting issues.
    
    \begin{table}[H]
    \caption{Survival prediction results. The multiple models for survival prediction were trained, and the best model \textbf{CoxPH} using clinical, CT radiomics, and deep learning features achieved a C-index of 0.82 in cross-validation.
} 
    \centering 
    \begin{tabular}{l c} 
    \hline 
    Survival Models   & C-index \\ [0.5ex] 
    \hline\hline 
    CoxPH Regression (clinical) & 0.70 \\ 
    CoxPH Regression (clinical + PET radiomics) & 0.67 \\
    CoxPH Regression (clinical + CT radiomics)  & 0.68 \\
    CoxPH Regression (clinical + PET/CT radiomics) & 0.72 \\
    CoxPH Regression (clinical + deep learning features) & 0.76 \\
    \textbf{CoxPH Regression (clinical + CT radiomics + deep learning features)} & \textbf{0.82} \\\hdashline
    Random Survival Regression (clinical) & 0.59 \\
    Random Survival Regression (clinical + PET radiomics) & 0.60 \\
    Random Survival Regression (clinical + CT radiomics) & 0.61 \\
    Random Survival Regression (clinical + PET/CT radiomics) & 0.59 \\
    Random Survival Regression (clinical + CT radiomics + deep learning features) & 0.58 \\\hdashline
    DeepSurv (clinical) & 0.60 \\
    DeepSurv (clinical + PET radiomics) & 0.68 \\
    DeepSurv (clinical + CT radiomics) & 0.69 \\
    DeepSurv (clinical + PET/CT radiomics) & 0.73 \\
    DeepSurv (clinical + PET/CT radiomics + deep learning features & 0.65 \\ [1ex] 
    \hline 
    \end{tabular}
    \label{table:surv} 
    \end{table}
    
    \newpage
    \section{Conclusion } 
    \label{conclusion}
    \indent We proposed a multimodal 3D H\&N tumor segmentation model, NormResSE-UNet3+, combining the squeeze-and-excitation layers~\cite{ref_article8} in a UNet3+~\cite{ref_article9} architecture to take advantage of full scale features allowing the model to  simultaneously focus on the relevant regions of interest. 
    The combination of both local information and global (e.g. context) information aimed to improve the accuracy of the segmentation. 
    We investigated the proposed neural network architecture with different training schemes and different loss functions (the Lovasz-Softmax loss and the Tversky loss~\cite{ref_article25}), however they did not significantly improve the overall segmentation performance when compared to the hybrid loss function of Log Cosh Dice~\cite{ref_article10} and Focal loss~\cite{ref_article11}, which was used in our final model.
    In turn, a method to post-process the predicted segmentation outputs based on uncertainty using Conditional Random Fields~\cite{ref_article12,ref_article13} to filter out false-positives and refine boundaries improved segmentation accuracy. 

    Future work will include Bayesian uncertainty measurements followed by a tailored post-processing technique based active-contour algorithms~\cite{ref_article26} for multimodal PET and CT images. 
    Graph-based methods and volume clustering~\cite{ref_article27} and multi-task learning~\cite{ref_article28} have been shown to improve segmentation tasks and will also be considered in future work.
    Increasing the multi-centre sample size for training and validation is expected to strengthen model inferences in order to demonstrate the robustness of the model and its ability to generalise. 
    A larger sample size would also enable it to make stronger inferences to improve the prediction of progression free survival. 
    Further work is required to reduce over-fitting issues  in progression free survival e.g. by adding regularization to the model. 
    The addition of mid-level deep learning features effectively improved progression free survival predictions compared to baseline models with only clinical and radiomic features both on training and test sets. 
    The extraction of relevant features is an active area of research and will be the focus of future work along with work on model architectures and custom loss functions. \\
    \\
    \textbf{Acknowledgment }\\
    
    This work was supported by the EPSRC grant number EP/S024093/1 and the Centre for Doctoral Training in Sustainable Approaches to Biomedical Science: Responsible and Reproducible Research (SABS: R³) Doctoral Training Centre, University of Oxford.
    The authors acknowledge the HECKTOR 2021 challenge for the free publicly
    available PET/CT images and clinical data used in this study~\cite{ref_article1}. 
    \newpage
    
    %
    %

    \bibliographystyle{splncs04}

\end{document}